\documentclass[preprint,showpacs]{revtex4}
\usepackage{bm}
\input{epsf}
\begin{document}
\title{Probing liquid surface waves,
liquid properties and liquid films with light diffraction}
\author{Tarun Kr. Barik, Partha Roy Chaudhuri, Anushree Roy and
       Sayan Kar }
\email{sayan@phy.iitkgp.ernet.in} \affiliation{Department of
Physics, Indian Institute of Technology, Kharagpur 721 302, India}
\begin{abstract}

Surface waves on liquids act as a dynamical phase grating for
incident light. In this article, we revisit the classical method of
probing such waves (wavelengths
of the order of mm) as well as inherent properties of liquids
and liquid films on liquids, using optical diffraction.
A combination of simulation and experiment is proposed
to trace out the surface wave profiles in various situations
(\emph{eg.} for one or more vertical, slightly immersed,
electrically driven exciters). Subsequently, the surface tension and
the spatial damping coefficient (related to viscosity) of a
variety of liquids are measured carefully in order to gauge the
efficiency of measuring liquid properties using this optical
probe. The final set of results deal with liquid films where
dispersion relations, surface and interface modes, interfacial tension
and related issues are investigated in some detail, both theoretically and
experimentally. On the whole, our observations and analysis seem
to support the claim that this simple, low--cost apparatus is
capable of providing a wealth of information on liquids and liquid
surface waves in a non--destructive way.

\end{abstract}
\maketitle
\def\d{{\mathrm{d}}}
\section{Introduction}
Optical probes are usually non--destructive -- they do not change
the properties of the medium being probed in any significant way
(unless the probe power is large). Diffraction and scattering of
light, as is known over centuries, are therefore capable of
providing a variety of information. In this article, we revisit
a well-known classical method of investigating liquid surface
waves using light diffraction. Although this method is
well--studied in the context of measurements of surface tension of
liquids \cite{weisbuch:1979}, our motivation for revisiting it is
primarily aimed at demonstrating its utility and capability in
understanding a wider class of situations/phenomena. In
particular, we investigate two aspects -- surface wave profiles
and liquid properties for waves on a single liquid and also for
waves on liquid films.

Before we begin with details, it is worth summarizing some of the
earlier work on liquid as well as solid surface waves. Use of
laser light scattering is a unique technique for non-contact,
noninvasive study of surface acoustic waves and capillary
waves \cite {wolf:1973}. Technological advances directed toward
diverse applications of surface acoustic waves on solid
surfaces are now well established, and has resulted in important
developments in the fabrication of practical devices \cite
{morgan:2000, Khan:1999, Tsai:1992, Deger:1998, Rimeika:2000,
phystoday:2002, Stegeman:1976, Monchalin:1986, duncan:2000}.

However, relatively few articles deal with surface waves on
liquid surfaces \cite{surface, earnshaw:1987, miao:2000,
barik:2005, klipstein:1996, Walkenhorst:1998, klemens:1984,
langevin:1992, ivanov:1988}. To mention a few, high frequency
capillary waves at the surface of liquid gallium and the mercury
liquid-vapor interface have been studied by means of quasi-elastic
light-scattering spectroscopy \cite {kolevzon:1996}. The spatial
damping coefficient of low-frequency surface waves at air-water
interfaces, using a novel heterodyne light--scattering technique,
has been measured and reported in \cite{lee:1993}. Quasi-elastic
light scattering has been used to study thermally excited
capillary waves on free liquid surfaces over a considerably wider
range of surface wavenumbers \cite{earnshaw:1987, klipstein:1996}.
Measurements of the decay coefficient of capillary waves in
liquids covered with monolayers of stearic acid, oleyl alcohol,
and hemicyanine are reported in \cite{saylor:2000}. Also, the
measurement of the pressure-area compression isotherm in
Langmuir monolayer films using the laser light diffraction from
surface capillary waves is reported in \cite{Martin:1993}.

In a recent article, we have shown how the effect of interfering
waves on a liquid surface could be inferred from the nature of the
diffraction pattern  \cite{barik:2005}. Here we begin by proposing
a combined method of simulation and experiment, sufficiently
general in nature and capable of verifying the nature of the
actual liquid surface wave profile. Having known the profile, we
then focus on the dispersion relation and discuss experiments to
determine the surface tension and viscosity of liquids. Finally,
we turn to liquid films where we try to understand the profile
through the various modes (surface and interface) and then go on
to check the values of interfacial tension.

Section II describes the basic scheme of our experimental
arrangement. The  methodology of our experiment and typical
results pertaining to the surface wave profile are discussed in
section III. In section IV, we discuss theory, observations and
measurements of surface tension and viscosity for different
liquids. Section V first discusses the theoretical platform for
surface capillary wave on liquid films. We then verify our theory with
experiments. Finally, in section VI, we summarize our results with some
concluding remarks.

\section{Experimental details and the theoretical background}

\subsection{Experimental setup}
\begin{figure}[htbp]
\centerline{\epsfxsize=3.25in\epsffile{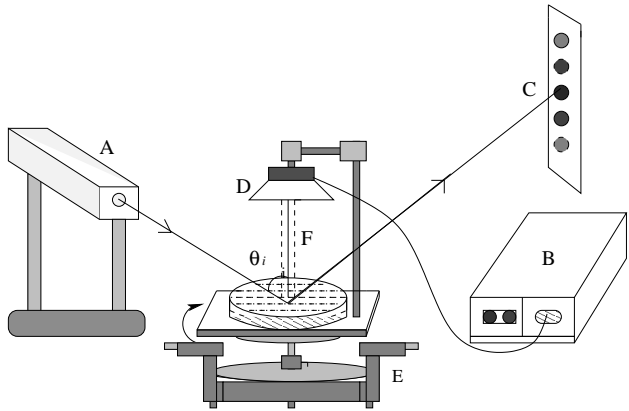}}
\caption{Schematic diagram of the experimental setup to study
capillary wave on liquid; A: Laser, B: Frequency generator, C:
Screen, D: Loudspeaker, E: Spectrometer and prism table assembly
and F: Exciter.} \label{setup}
\end{figure}

 As shown in Fig. \ref{setup}, a petridish of diameter about 18.5 cm is filled
 with the experimental liquid to a depth of $\sim$ 1 cm. A metal pin
 with its blunt end glued vertically upright to the diaphragm of a
 loud speaker (held above the petridish) acts as an exciter. When slightly immersed in the
 liquid and driven by a low frequency sinusoidal signal generator,
 this exciter vibrates vertically up and down and generates the
 desired liquid
 surface waves. To study the profile of surface waves we place the
 petridish-loud-speaker assembly on the prism-table of a spectrometer in
 order to choose the diffracting element at any position on the liquid surface
 and to record the corresponding position from the prism-table scale (with a least count of 0.1 degree). Light
 from a 5 mW He-Ne laser ($\lambda$ = 632.8 nm) having a beam diameter of
 $\sim$ 1.8 mm is directed to fall on the liquid surface at an angle of
 incidence of $77^{0}$ ($\theta_i$), as noted in our experiment. The laser beam
 incident on  the dynamical phase grating formed on the liquid
 surface, produces Fraunhofer diffraction pattern which is observed
 on a screen placed at a fairly large distance (3.42 meters in our
 case) from the diffraction center. The images of the diffraction
 pattern are recorded using a digital camera (make Sony, DSC - P93, 5.1
 mega pixels). Spurious noise in the image is removed using Photoshop
 software. The intensity across the diffraction spots are measured
 using a photo-detector operating in the linear dynamic range of the
 photo-currents' response to optical power. During our experiment, the
 room temperature recorded is $ 25^{0}$C.

\subsection{Theoretical background}

We discuss here, briefly, the diffraction mechanism which lies at
the heart of all our experiments. 
When monochromatic light of wavelength $\lambda$ is incident on a
circular surface wave of frequency $\omega$ and wavevector
$K$, the phase modulation produced by the surface wave is given as
\begin{equation}
\phi (x^{\prime}) = \frac{2\pi}{\lambda} \left [ (2 h \cos
\theta_{i}) \sin \left ( \omega t - \frac{Kx^{\prime}}{\cos
\theta_{i}} \right ) \right]
\end{equation}
where, the factor of $\cos \theta_{i}$ appears due to oblique
incidence of the incident monochromatic light \cite{duncan:2000, miao:2000, 
barik:2005, goodman:1968}.
The field strength $E$ of the diffraction pattern can be estimated
from the Fourier transform of the aperture function (in this case
the surface wave produced by single exciter). The intensity of the
diffraction patterns, obtained from $EE^{*}$, is given by
\cite{goodman:1968, hecht:2003}
\begin{equation}
I(x^{\prime}) = \sum_n J_n^2 \left (4\pi h \cos \theta_{i}/
\lambda \right ) \delta \left (\frac{x^{\prime}}{\lambda z} -
\frac{n}{\Lambda \cos \theta_{i}}\right )
\end{equation}

\noindent where $z$ is the horizontal distance between the
location of the laser spot on the liquid surface and the screen
and $\Lambda= \frac{2\pi}{K}$, is the wavelength of the surface
wave. $h$ is the amplitude of the surface wave. $x^{\prime}$ is
the coordinate which measures the distance of the diffraction
spots from a reference point (central spot) on the observation
plane. $J_n$ is the Bessel function of order $n$ and $\delta ()$
denotes the Dirac delta function. The intensity distribution on
the diffraction pattern vanishes at points where $J_n$ is zero
satisfying Eqn. 2. The quantity involving the Bessel function in
this equation gives the magnitude of the intensity of diffraction
spots for different orders $n$, while the positions of the spots
are given by the delta function.
\begin{figure}[htbp]
\centerline{\epsfxsize=3.25in\epsffile{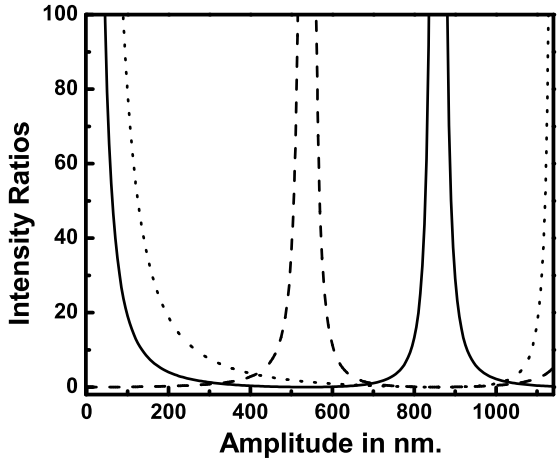}}
\caption{Theoretical plot of wave amplitude versus intensity ratio
for $\theta_{i}$ = 77 $^{0}$, $\lambda$ = 632.8 nm. Solid line,
dashed line and dotted line stand for the ratios  $I_{0}/I_{1}$,
$I_{1}/I_{0}$ and $I_{0}/I_{2}$ respectively.} \label{amplitude}
\end{figure}

When we generate capillary waves by means of a single pin exciter,
the amplitude of the wave at the point of oscillation is almost
the same as the amplitude of the oscillating pin in a low viscous
liquid \cite{barik:2005}. The intensities of different orders
depend only on the amplitude of the capillary wave ($h$), if other
parameters like $\theta_{i}$ and $\lambda$ are kept constant (Eqn.
2). Theoretical plots of wave amplitude versus the intensity
ratios, $I_{0}/I_{1}$, $I_{1}/I_{0}$ and $I_{0}/I_{2}$, using Eqn.
2 (keeping $\theta_{i}$ and $\lambda$ fixed) are shown in Fig.
\ref {amplitude}. In our experiment, for a particular $\omega$,
the intensity of different order spots in the diffraction pattern are measured
by a photo-diode detector when the laser beam is focused along a
radial line on the liquid surface. The intensity ratios like
$I_{0}/I_{1}$ and $I_{0}/I_{2}$ are then evaluated. By comparing
our experimentally measured ratios with those obtained from the
theoretical plot (Fig. \ref{amplitude}), we estimate the average
value of $h$. It is to be noted that, for $I_{0} << I_{1}$, the
value of this ratio $I_{0}/I_{1}$ becomes very small and could
yield erroneous results. In such cases, we used the ratio
$I_{1}/I_{0}$ for estimation of $h$ (however, we were always below
the limit for which this ratio becomes very large). Again, by measuring
$x^\prime$,  we determined the wavelength ($\Lambda$) of the surface 
capillary waves to be 2.1 mm [for more details see \cite{barik:2005}].
For several exciters placed at different types of geometric configurations 
(eg, along a line, or a polygon etc.), the surface wave patterns change, and
hence Eqn. 2 must also change.  

\section{The Surface Wave Profile}
\subsection{Simulation of the profile}

\begin{figure}[htbp]
\centerline{\epsfxsize=6.5in\epsffile{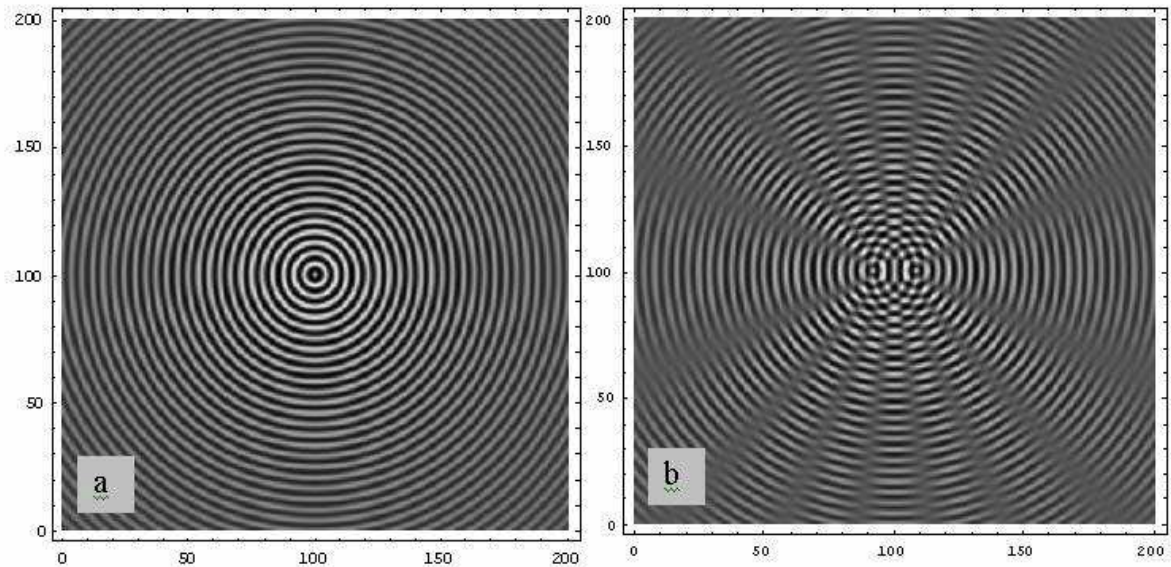}} \caption{The
simulated surface capillary wave profiles for (a) single exciter
and (b) two exciters.}
\label{sscwp}
\end{figure}

Low frequency surface waves are not visible to the naked eye. One
may appreciate this when we simulate the surface wave profile
using realistic dimensions and parameters. In a previous study
\cite{barik:2005}, we measured the surface-wave parameters
(wavelength, amplitude at the point of excitation)  at a given
oscillation frequency (220 Hz). Here, we use similar typical
values to generate theoretically the distribution of waves on the
liquid surface. The equations, which we use to simulate the wave
profile for the cases of single and  double-pin exciters are (at a
fixed time `$t$', say $t$=0)

\begin{equation}
 \psi_1 (X,Y) = h\exp \left ( -\delta\sqrt{\left (X-x_1\right)^2+\left
  (Y-y_1\right)^2}\right)\cos \left  (\frac{2\pi}{\Lambda}\sqrt{
\left (X-x_1\right)^2+\left(Y-y_1\right)^2} \right)
\end{equation}
 and
\begin{eqnarray}
\psi_2 (X,Y) = h\exp \left ( -\delta\sqrt{\left
(X-x_1\right)^2+\left
  (Y-y_1\right)^2}\right)\cos \left  (\frac{2\pi}{\Lambda}\sqrt{
\left (X-x_1\right)^2+\left(Y-y_1\right)^2} \right)
+ \nonumber \\
  h\exp \left ( -\delta\sqrt{\left (X-x_2\right)^2+\left
  (Y-y_2\right)^2}\right)\cos \left  (\frac{2\pi}{\Lambda}\sqrt{
\left (X-x_2\right)^2+\left(Y-y_2\right)^2} \right),
\end{eqnarray}

\noindent
 respectively. Here $\left( x_1, y_1\right)$ and $ \left( x_2, y_2\right)$ are
the centers of oscillations and $\delta$ is the spatial damping
coefficient of the liquid. To mimic the intended experiments, we
assume that each wave has the same frequency $\omega$, wavelength
$\Lambda$, and amplitude  $h$ (at the center of oscillation).
In our simulations, we  use the typical values of $\Lambda$
= 2.1 mm, $h$ =1.0 mm and $D$ = 8.4 mm (separation between
pin-exciters) \cite{barik:2005}. $\delta$ is chosen to be 0.235
cm$^{-1}$ (for water at 220 Hz, we obtain this value from
experiments discussed later in section IV). The corresponding
surface-wave profiles estimated for the two cases of interest are
shown in Fig. \ref{sscwp}(a) and Fig. \ref{sscwp}(b),
respectively. Following this recipe one can obtain the
surface-wave profile for any number of oscillation sources. The
simulated profiles enable a better visualization, which, in turn,
can act as a guideline for their actual experimental verification.

\begin{figure}[htbp]
\centerline{\epsfxsize=3.5in\epsffile{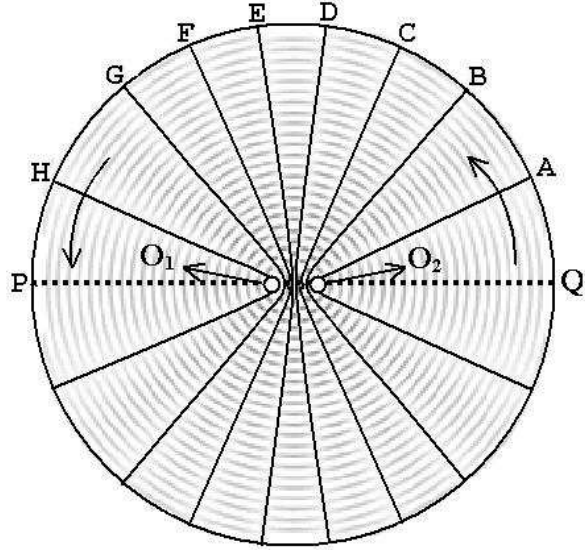}} \caption{The
schematic of our experiment, which we have followed to verify the
simulated two source pattern. O$_1$ and O$_2$ are the two
exciters. Solid lines are the hyperbolic loci along the direction
of destructive interference.}
\label{schem}
\end{figure}

\subsection{Experimental results for two source pattern}
\begin{figure}
\centerline{\epsfxsize=3.25in\epsffile{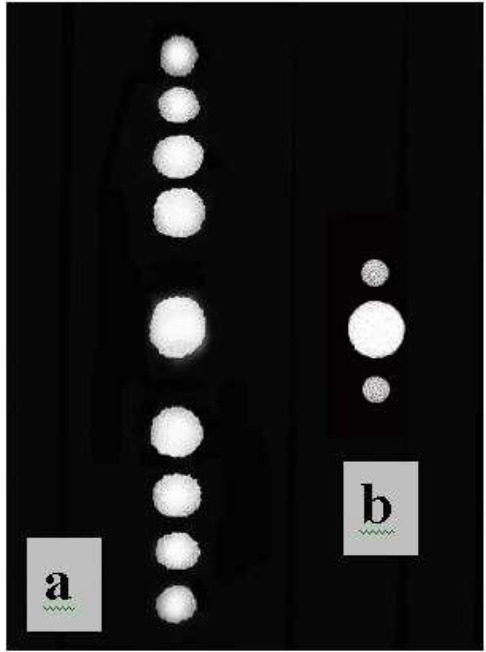}}
\caption{Diffraction patterns when light is incident along the
  direction of  (a) constructive interference and (b) destructive interference
for the two source (exciter) case.}
\label{pattern}
\end{figure}

The simulated profiles are built on the assumptions made in Eqn. 3
and Eqn. 4. It is necessary to verify experimentally whether the
assumptions are good enough to represent the characteristics of
the surface waves. We note that the simulated  two source
interference pattern [shown in Fig. \ref{sscwp}(b)] exhibits
hyperbolic loci for the constructive and destructive interference
nodes. Using this fact as a guideline, we trace the hyperbolic
loci (maxima and minima of the interfering waves) on the liquid
surface by observing the changes in the corresponding
 diffraction patterns. Initially, when the laser beam is incident along
 the central line PQ (see Fig. \ref{schem}) where constructive interference of
 waves occur,
 the diffracted light shows, as  expected \cite{barik:2005}, a central spot
 along with higher order ones symmetrically located on either side of
 the central spot [see Fig. \ref{pattern}(a)]. As we shift the probe beam
 gradually away from
 the central line along a circle (note the arrow in Fig. \ref{schem}), the number of higher order spots
 progressively decreases till we reach the position of the adjacent hyperbola (O$_2$A), where the diffraction should
 ideally contain only the
 central order. In experiment, however, at this position, we
 find the first order spots too, though with very low intensity [see
 Fig. \ref{pattern}(b)]. This
 happens due to the finite spot size of the laser beam. Nevertheless, beginning with the central line at
one side and gradually rotating the prism-table till we reach
$180^{0}$, we observe a repetition of the patterns [shown in Fig.
\ref{pattern}(a) and \ref{pattern}(b)] symmetrically at all complementary
angles. We note the
 angular positions of all consecutive minima. These results are shown in Table-I. Thus, a comparison of our
experiment (Fig. \ref{schem}) with the simulation [upper half of
Fig. \ref{sscwp}(b)] for identical values of  parameters shows the
same number (eight) of destructive interference lines at almost
the same angular separation. Beyond $180^{0}$ and upto $360^{0}$,
($i.e$, while scanning the lower semicircular region in Fig.
\ref{schem}) identical patterns are observed. This confirms that
the Eqn. 4 used to describe and simulate the two source surface
wave profile is, indeed, a correct representation of the observed
profile.
We note that our method of investigation is
sufficiently general (we have indeed checked upto six exciters)
and can be used for several exciters located along a line or in
other geometric configurations (eg. exciters on a triangle, quadrilateral, pentagon, hexagon etc.).

\subsection{Minima distribution for oscillation sources in regular array}

\begin{figure}
\centerline{\epsfxsize=6.5in\epsffile{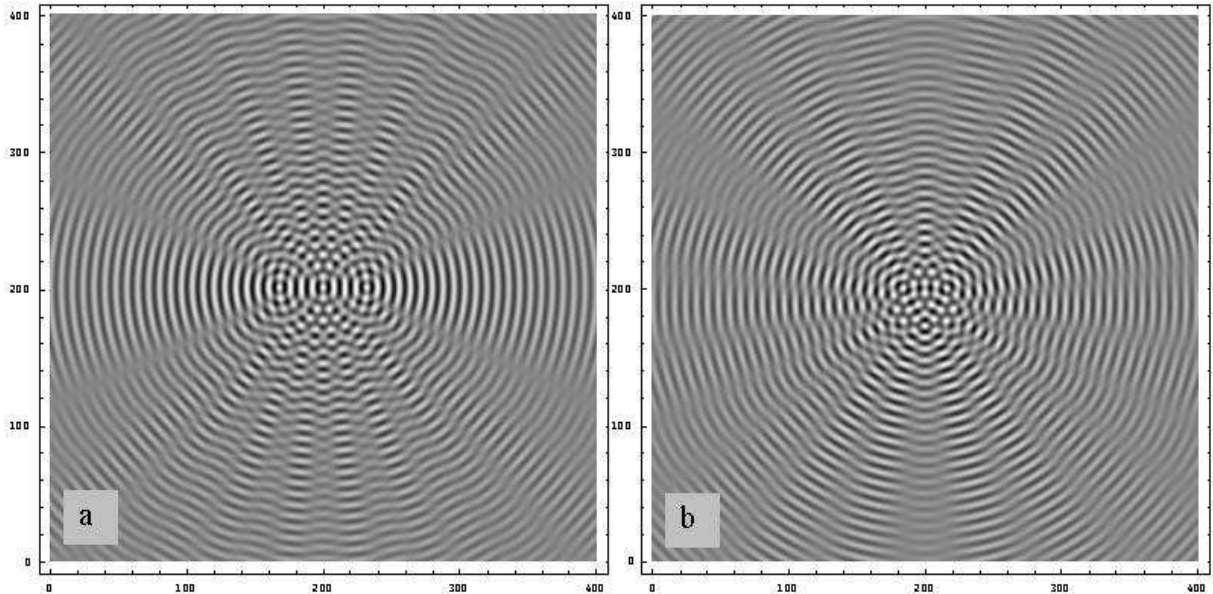}}
\caption{Simulated interference profile of surface capillary wave for (a)
three exciters in a straight line and (b) three exciters at the vertices of
an equilateral triangle.}
\label{profile}
\end{figure}

When multiple oscillating sources are used, different surface wave
profiles can be generated by appropriately configuring their
positions. It is interesting to study these interference patterns
of the surface waves for multiple sources (e.g., a linear chain of
oscillation sources or those placed in a closed regular geometric
figure). A typical example of a simulated interference pattern for
three oscillation sources is shown in Fig. \ref{profile}. For this
purpose, we choose to trace the distribution of minima lines in
the simulated interference pattern in terms of the number of lines
and their angular position. It is obvious that the number of
minima lines increases with the increase in the number of
oscillation sources. For example, our simulation of the pattern
for 2, 3, 4, 5 and 6 oscillation sources placed in a straight
line with a spacing of $D = 8.4$ mm between the adjacent two
oscillators, and $\Lambda = 2.1$ mm, exhibits the number of
hyperbolic lines respectively as 8, 16, 24, 32, and 40. To derive
this behavior theoretically, we use the  interference treatment for
n-oscillating sources. We consider n ($\geq 2$) sources lying in a
straight line. Representing the wave generated by each oscillating
source as $h\exp(i({\bf K.r_n}-\omega t))$, where $h$ is the
amplitude and ${\bf r_n}$ is the position vector from the n-th
source to the point of observation, we find the resultant field
$\psi_P$ at a certain point P as \cite{hecht:2003},

\begin{equation}
\psi_p=h\frac{Sin(\frac{n\beta}{2})}{Sin(\frac{\beta}{2})}\exp \left(i(KR-\omega t)\right),
\end{equation}

\noindent
where $R=\frac{1}{2}D(n-1)Cos(\phi)+\bf r_1$, is the distance from the
centre of the line of oscillators to the point P and $\beta=KDCos(\phi)$, is
the phase difference between adjacent sources. Again $\phi$ is the angle
between $R$ and the line joining the sources of oscillation. For minima at P,
the condition is $h\frac{Sin(\frac{n\beta}{2})}{Sin(\frac{\beta}{2})}=0$.
Using the boundary condition mentioned in \cite{barik:2005}, we calculate the allowed number of minima lines and their corresponding angular positions. For $n=3$, for example, the above minima condition becomes $\frac{D}{\Lambda}Cos(\phi)=\pm(\frac{m}{2})$, where m is an integer but not equal to or multiple of $n$. If we choose $\frac{D}{\Lambda}=4$, then $Cos(\phi)$ has only 16 allowed values, i.e. 16 hyperbolic minima lines are
possible. The angular ditribution of these asymptotic lines perfectly match
with the observed pattern. We have carried out this comparison successfully for higher order $n$-values upto 5. From the above calculations we have seen that the
number ($N_o$) of asymptotic minima lines in the region above (or below) the
line joining the oscillation sources is directly proportional to the ratio
$\frac{D}{\Lambda}$ and also the number of the interval ($n-1$) between the exciters. Taking care of the $\pm$ sign of $Cos(\phi)$, we can express $N_o$ as

\begin{equation}
N_o=\frac{2D}{\Lambda}\left(n-1\right),
\end{equation}
 which gives the same result (for different n nalues) as obtained before. This relationship can, just by measuring the number of minima lines for given $D$ and $n$, estimate $\Lambda$ of the capillary waves.

Next we study n ($\geq 3$) oscillation sources forming a closed loop shape
of an equilateral triangle, square, pentagon, and a hexagon. The number ($N_c$) of minima lines (which are lying on ether side of a symmetric axis) in this case can be similarly expressed as,

\begin{equation}
N_c=\frac{2D}{\Lambda}(\frac{n}{2}),
\end{equation}

The number of lines for the above stated closed geometrical figures at the
same values of $D$ and $\Lambda$ are 12, 16, 20 and 24 respectively. We have
not been able to calculate exactly the angular distribution of the minima
lines. However, using the simulation mentioned above, one can determine the
interference pattern due to any number of oscillation sources placed
along a line or at the vertices of polygons (regular or irregular). Such a study might be useful for generating a desired dynamic phase grating aperture for 
diffraction based micro-photonic systems.

\section{The dispersion relation and liquid properties using light diffraction}

Knowing the surface wave profile is, of course, not enough. A
crucial element of any wave phenomenon is the dispersion relation,
which, obviously, contains within it, quantifiers of material
properties (here, liquid properties, such as surface tension and
viscosity). In this section we describe how we use Fraunhofer
diffraction of laser light by the surface profile of ripples on
liquids, described in Section III, to study such liquid
properties. Several methods to accurately determine these two
properties are now well-established and available in the
literature. For measurement of surface tension, techniques widely
used are the Capillary Rise method, Drop-weight method, Jaeger's
method, Rayleigh's method etc. \cite{osipow:1993,Ghosh:1985}.
Likewise, for the measurement of viscosity, popular methods like
Poiseuille's method, Stokes' method etc., are well-known and gives
accurate results \cite{osipow:1993,Ghosh:1985}. Here, we focus on
a single optical set up which can measure both surface tension and
viscosity of low-viscous liquids with reasonable accuracy. Though
the measurement of surface tension of liquids using Fraunhofer
diffraction of laser light by surface waves is well known in the
literature \cite{weisbuch:1979}, to the best of our knowledge the
estimation of the viscosity of a liquid using this probe and, in
particular,  the simple method which we have followed, is
certainly new. In the following we elaborate on the above
mentioned measurements in some detail.

\subsection{Surface tension}

Consider the well-known dispersion relation for surface capillary wave
on liquids
\cite{levich:1962,landau:1975,kinsler:1982}
\begin{equation}
\omega^{2} = \left (gK + {\alpha} K^{3}/\rho\right ).
\end{equation}

\noindent Here $\omega$ and ${K}$ are the angular frequency and
wavevector of the surface capillary wave, respectively. $\alpha$ and $\rho$
are the surface tension and density of the liquid and $g$ is the
acceleration due to gravity. The above relation shows that waves
at the surface of the liquid are dominated by gravity as well as
surface tension. Neglecting the gravitational effect and taking
the logarithm on both sides of the Eqn. 8 we get

\begin{equation}
\ln{\omega} = \frac{3}{2}\ln{K} + \frac{1}{2}\ln{\left
(\frac{\alpha}{\rho} \right )},
\end{equation}

\noindent which is an equation of a straight line for
$\ln{\omega}$ \emph{vs.} $\ln{K}$ with a slope of $\frac{3}{2}$
and y-intercept $ \frac{1}{2}\ln{\left (\frac{\alpha}{\rho} \right
)}$, which contains the surface tension $\alpha$.

\begin{figure}
\centerline{\epsfxsize=4.25in\epsffile{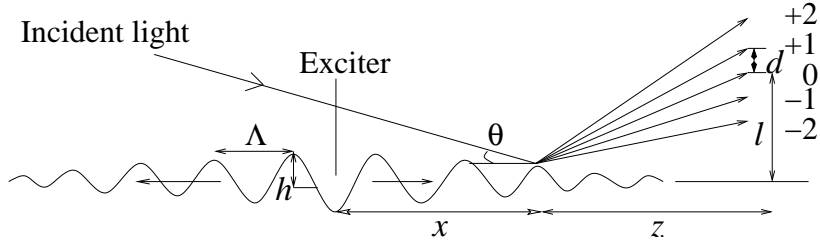}}
\caption{Geometrical representation of the basic measurement on
surface tension and viscosity.} \label{visg}
\end{figure}

In our experiments, we have generated the surface capillary
wave on the liquid surface with a single exciter. As we discussed
before, the expected surface wave profile [shown in Fig.
\ref{sscwp}(a)] acts as a reflection phase grating for the
incident laser light and the intensity distribution satisfies Eqn.
2. If we take the separation between the central order and the
first order ($n$=1) diffraction spots to be $d$ (see Fig.
\ref{visg}), then from the  delta function in Eqn. 2 we get the
expression for the wavenumber $K$ as

\begin{equation}
K = \frac{2 \pi d}{\lambda z} \mbox{sin}{\theta},
\end{equation}

\noindent
where  $\theta = \frac{\pi}{2}-\theta_{i}$ is the
grazing angle of incidence of the laser beam. For small grazing
angle, we can approximate $K$ as
\begin{equation}
K = \frac{2\pi ld}{\lambda z^2},
\end{equation}

\noindent where $l$ is the perpendicular height of the central
order spot from the liquid surface level (see Fig. \ref{visg}). As
$\omega$ changes, the distance between bright spots of the
diffraction pattern also changes. Sets of diffraction patterns are
obtained on the screen for different values of $\omega$. For each
set, we have traced the positions of spots of different order on
the screen.

For Fraunhofer diffraction, values of $z$ and $l$ in Eqn. 11 are
quite large (342 cm and 87 cm, respectively, in our experiment).
Thus, the systematic error introduced in measuring these lengths
with a metre  scale (with a least count of 1.0 mm) is quite less.
On the other hand, a careful and accurate measurement of $d$ (of
the order of few mm) is crucial for evaluating $K$. From the
characteristic feature of the delta function in Eqn. 2, we find
that the diffraction spots due to different orders are equidistant
for a particular frequency of capillary wave. Thus, to reduce the
systematic error in measuring $d$, we have measured the separation
between the positive and the negative fourth-order (fifth-order)
spots. The values of $d$ are obtained through averaging these
data. For each frequency $\omega$, we have repeated the experiment
at least four times. Hence, using Eqn. 11, we estimate the value
of $K$ for a particular $\omega$. In addition, it is reasonable to
assume that the frequency, $\omega$, of capillary wave is the same
as that of the exciter (driven by a function generator)
\cite{barik:2005}. Thus, the values of $\omega$ are found directly
from the frequency readings of the function generator driving the
exciter. Following these measurements, we plot a ln -ln  graph for
$K$ \emph{vs.} $\omega$ and evaluate the slope and y-intercept of
this curve. The slope of the curve ($\frac{3}{2}$) should verify
the power law predicted by the dispersion relation while the
y-intercept should give the surface tension.

\begin{figure}
\centerline{\epsfxsize=4.5in\epsffile{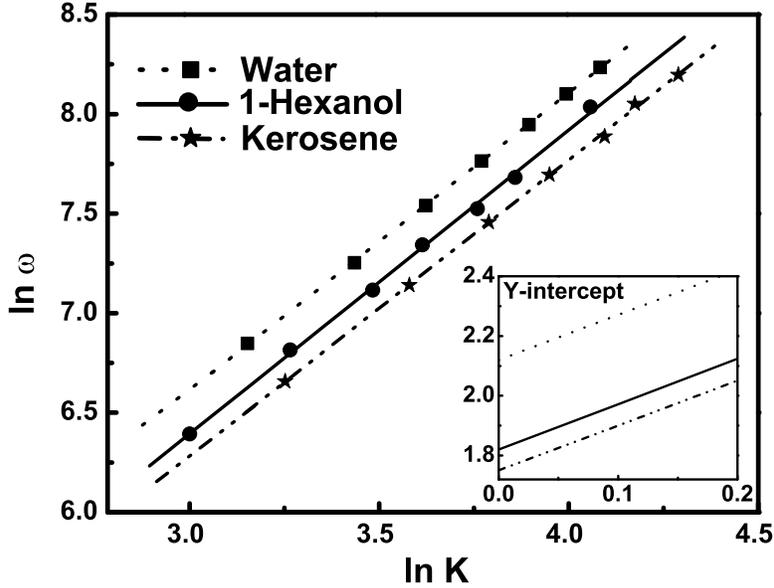}} \caption{ The
ln-ln plots of the wavevector (K) versus the frequency ($\omega$)
exhibit a straight line of slope very close to $\frac{3}{2}$
obtained for three different liquids. The inset shows the
corresponding y-intercepts.}
\label{surface}
\end{figure}

Our experimental results on three different liquids (water,
1-hexanol and kerosene -- a mixture of petrolium hydrocarbons) are
shown in Fig. \ref{surface}. The non-linear least square fit
(shown by lines in Fig. \ref{surface}) to the data with Eqn. 9
verifies the 1.5 exponent of the dispersion relation within
experimental uncertainty (1.5$\pm$0.03). Using the known value of
$\rho$ for each liquid, we estimate the value of $\alpha$ from the
measured y-intercept of the straight lines. It is to be noted that
the maximum percentage error in measuring $\alpha$ is limited to
6\% - 8\% (for different liquids) by our measurement procedure. We
have tabulated the measured values of surface tension for
different liquids in Table-II. These results match quite well with
the standard values of surface tension for the corresponding
liquids \cite{Lide:1998}.

\subsection{Viscosity}

\begin{figure}[htbp]
\centerline{\epsfxsize=3.25in\epsffile{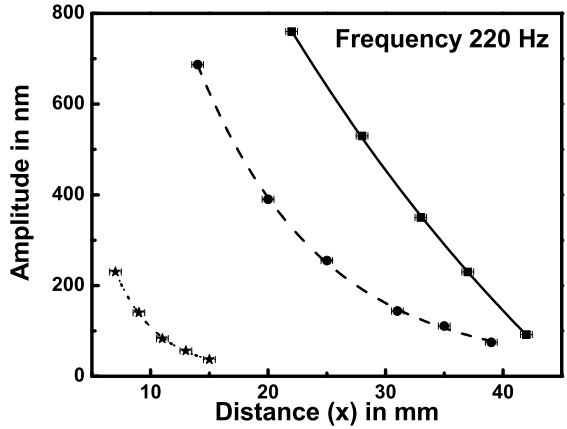}}
\caption{Experimental plot of distance ($x$) versus surface
capillary wave amplitude ($h$). Solid line, dashed line, and dotted
line for the liquids water, kerosene, and 1-hexanol respectively.}
\label{viscosity}
\end{figure}

Using the same experimental set-up, we now move on towards
estimating the viscosity of a given liquid.  We mentioned before
that for capillary waves generated by means of a single pin exciter, the
amplitude of the wave at the point of oscillation is almost the
same as the amplitude of the oscillating pin in a low viscous
liquid \cite{barik:2005}. However, if we imagine a radial straight
line from the source of excitation (on the liquid surface), then
along this line the amplitude of surface wave is gradually damped
due to the viscosity of the liquid (see Fig. \ref{visg}). The
laser beam is focused along a radial line at different distances
on the liquid surface from the point of excitation ($x$ in Fig.
\ref{visg}). We have seen in Sec. II that the intensities of the
diffraction spots of different orders depend only on the amplitude
of the capillary wave, if other parameters like $\theta_{i}$ and
$\lambda$ are kept constant (Eqn. 2). By comparing our
experimentally measured intensity ratios of different orders in
the diffraction pattern with those obtained from the theoretical
plot (Fig. \ref{amplitude}), we estimate the average value of $h$
at a particular location on the liquid surface. In Fig.
\ref{viscosity}, we plot $x$, the distance from the point of
oscillation \emph{vs.} $h$, the capillary wave amplitude for three
experimental liquids (water, kerosene, and 1-hexanol). We fit the
data points with an exponential decay function, where we kept the
decay constant (also known as spatial damping coefficient of the
liquid), $\delta$, as varying parameter. The values of $\delta$
for three different liquids  are tabulated in Table-II.

The wavevector for a wave of fixed real frequency, $\omega$, has a
small imaginary part, which contributes to the spatial damping.
Substituting $K=K_{0}+i\delta$ (where $\delta << K_0$) in the
well- known Navier-Stokes equation for waves on a liquid-air
interface, one gets\cite{taylor:1978,lee:1993},

\begin{equation}
\delta = \frac{4\eta\omega}{3\alpha},
\end{equation}

\noindent where $\eta$ is the viscosity of the liquid.
Experimentally, we have obtained the values of $\alpha$ and
$\delta$ for the above mentioned liquids. The estimated value of
viscosity of different liquids, using Eqn. 12, are shown in
Table-II. The experimental plots in Fig. \ref{viscosity} are for
frequency 220 Hz. We have also checked our results for two other
frequencies (260 Hz and 300 Hz) and have obtained similar results
(see Table-II).

\section{Liquid films on liquids: theory and experiments}

Till now we have been exclusively concerned with waves on the
surface of a single liquid, or, more precisely, waves at the liquid-air
interface. A more complicated scenario arises when there exists a film of
a different liquid on top of a given liquid (the two liquids are
immiscible). In such a situation, the dispersion relation changes
drastically, which, in turn, changes the surface wave profile. We now
investigate the novelties that arise through studies with our optical
probe.

\subsection{Background theoretical framework}

The dispersion relation in Eqn. 8 for capillary wave propagation
on liquid surfaces is valid only when the depth of experimental
liquid is large. For lower values of the depth of the liquid
(denoted by, say $y$), the relation becomes \cite{landau:1975}
\begin{equation}
\omega^{2} = \frac{\alpha K^{3}}{\rho}\tanh \left ( K y\right)
\end{equation}
\noindent where, here too, we neglect a term due to gravity.
Typically, if we choose $y = 1$ cm and a range of $K$ from 20 to
80 cm $^{-1}$ (as used in our earlier experiment), $\tanh\left(K
y\right)$ is nearly unity, and the dispersion relation reduces to
the one given by Eqn. 8. The interesting problem we focus on now
is that of wave propagation on a liquid film spread out on another
liquid. If the thickness of the film is very low, then not only
the surface tension but also the interfacial tension across the
liquid-film boundary, play a role in the propagation mechanism of
the waves. We first work out a general dispersion relation where
the surface and interfacial tensions as well as the film thickness
appear explicitly in our mathematical expressions.

\begin{figure}[htbp]
\centerline{\epsfxsize=3.25in\epsffile{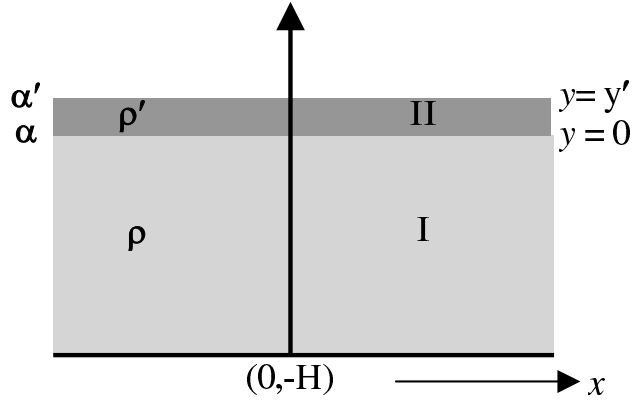}}
\caption{Liquid film on liquid and the various parameters used.}
\label{film}
\end{figure}

Fig. \ref{film} shows the symbols used to represent various parameters in
the following discussion. We denote the density of the lower
liquid by $\rho$ while $\alpha$ is the interfacial tension at the
liquid-liquid interface. The same quantities for the film-air
interface are represented with primes i.e., $\rho^{'}$ and
$\alpha^{'}$ respectively. The equilibrium plane of separation
(interface) between the liquids is at $y=0$. The liquid below the
$y=0$ plane extends upto a value $y=-H$ with $H$ being very large.
The film is located above $y=0$ with a thickness $y' << H$. Under
these assumptions, the general expressions for the surface
capillary waves in the region (I): -H $\leq y \leq 0$ and in region
II : $0 \leq y \leq y'$ are given by \cite{landau:1975},
\begin{equation}
\psi = A e^{Ky}\cos \left ( Kx-\omega t \right)
\end{equation}
and
\begin{equation}
\psi^{'} = \left [B e^{-Ky}+ C e^{Ky}\right]\cos \left ( Kx-\omega
t \right),
\end{equation}
\noindent
 respectively. Here A, B and C are arbitrary constants.
The above two waves must obey the following boundary conditions
arising from the continuity of the velocity component($v_{y}$) and
that of pressure at the surface of separation of the two liquids,
i.e.,
\begin{equation}
\frac{\partial\psi}{\partial y} = \frac{\partial\psi^{'}}{\partial y},
\end{equation}
\begin{equation}
g\rho\frac{\partial\psi}{\partial y} + \rho
\frac{\partial^{2}\psi}{\partial t^{2}}-\alpha \frac
{\partial}{\partial y}\left (\frac{\partial^{2}\psi}{\partial
x^{2}}\right) =\rho^{'} \left
(\frac{\partial^{2}\psi^{'}}{\partial t^{2}}\right )+g\rho^{'}
\left (\frac{\partial\psi^{'}}{\partial y}\right ),
\end{equation}
\begin{equation}
\rho^{'}g\frac{\partial \psi^{'}}{\partial y} +
\rho^{'}\frac{\partial^{2} \psi^{'}}{\partial
t^{2}}-\alpha^{'}\frac {\partial}{\partial y}\left (
\frac{\partial^{2}\psi^{'}}{\partial x^{2}}\right) = 0.
\end{equation}
\noindent Evidently, the first two conditions apply at $y=0$ while
the third one is valid at the top surface at $y=y^{'}$. In the
boundary conditions above, we neglect the effect of gravity in all
calculations below. Using Eqn. 14 and Eqn. 15 in Eqn. 16 and Eqn.
17 we obtain
\begin{equation}
A = C - B
\end{equation}
and
\begin{equation}
C = B \frac{\left [ \alpha K^{3}-\left( \rho +
\rho^{'}\right)\omega^{2}\right]}{\left [ \alpha K^{3}-\left(\rho
- \rho^{'}\right)\omega^{2}\right]}.
\end{equation}
\noindent Substituting Eqn. 19 and Eqn. 20 in Eqn. 18 with
$y=y^{'}$, we construct a quadratic equation for $\omega^{2}$
given by:
\begin{equation}
\omega^{4}\left[
1+r\tanh\left(Ky^{'}\right)\right]-\omega^{2}\left[
\omega_{s}^{2}\tanh\left(Ky^{'}\right)+r\omega_{s}^{2}+\omega_{i}^{2}
\right]+\omega_{s}^{2}\omega_{i}^{2}\tanh\left(Ky^{'}\right) = 0,
\end{equation}
\noindent where, we have used the notations $\frac{\rho^{'}}{\rho}
= r$, $\frac {\alpha^{'}K^{3}}{\rho^{'}}= \omega_{s}^{2}$ and
$\frac {\alpha K^{3}}{\rho}= \omega_{i}^{2}$ for simplicity. The
two roots of Eqn. 21 are the new dispersion relations for our
given problem. The two roots are
\begin{equation}
\omega_{\pm}^{2}= \frac{P\pm \sqrt{G} }{2Q},
\end{equation}
\noindent
where, $P=r\omega_{s}^{2}+\omega_{s}^{2}\tanh\left(Ky^{'}\right)+
\omega_{i}^{2}$, $Q=1+r\tanh\left(Ky^{'}\right)$, and $G=P^{2}-
4Q\omega_{s}^{2}\omega_{i}^{2}\tanh\left(Ky^{'}\right)$.

Consider two limiting cases of the above dispersion relation. Let
us first assume $y^{'}\rightarrow \infty$. Then
$\omega_{+}^{2}\rightarrow\omega_{s}^{2}$ and
$\omega_{-}^{2}\rightarrow\frac{\omega_{i}^{2}}{1+r}$ =
$\frac{\alpha K^{3}}{\rho + \rho^{'}}$. Thus, when $y'$ is very
large, the $\omega_{+}$ mode corresponds to the waves
propagating on the upper liquid surface (thus called surface
mode). On the other hand, the $\omega_{-}$ mode involves the
interfacial tension \cite{landau:1975} as well as the density of
the lower liquid and may be termed as an interface mode. In such a
situation ($y'$ large), reflecting the probe laser beam off the
surface wave profile on the liquid film, we can obtain information
(through the diffraction pattern) on wave propagation and
properties of the liquid film. The existence of the other mode can be
confirmed (for $y'$ large) only if we perform experiments in
transmitted light, though it is not clear how such information may
be obtained. On the contrary, assuming $y^{'}\rightarrow 0$ (very
thin film), we get
$\omega_{+}^{2}\rightarrow\left(r\omega_{s}^{2}+
\omega_{i}^{2}\right)$ = $\frac{
\left(\alpha^{'}+\alpha\right)K^{3}} {\rho}$, and
$\omega_{-}^{2}\rightarrow 0$. Thus the $\omega_{+}$ mode now
depends on properties of both the liquids, ($\alpha^{\prime}$,
$\rho$), as well as properties of the interface ($\alpha$).

Furthermore, we can obtain the constants $A$, $B$, $C$ and hence $\psi$
and $\psi^{'}$ by assuming a value for the amplitude imparted initially
at the top surface. In this way, one may obtain the amplitudes of $\psi$ and $\psi^{'}$ for both the modes $\omega_+$ and $\omega_-$. In particular, we have
noted (not demonstrated here) that the nature of variation of the
amplitudes of $\psi^{'}$ (with the film-hight $0<y'<h$)
for the $\omega_{+}$ mode is quite different from that for the
$\omega_-$ mode. In the former case ($\omega_+$) the value of the
amplitude drops to a much smaller value at the interface compared to
the drop in the amplitude for the $\omega_-$ mode. However, it is
not possible for us to check this fact through our experiments.
Hence we refrain from discussing this aspect further in this
article.

\subsection{Experiments and results}

\begin{figure}[htbp]
\centerline{\epsfxsize=5.6in\epsffile{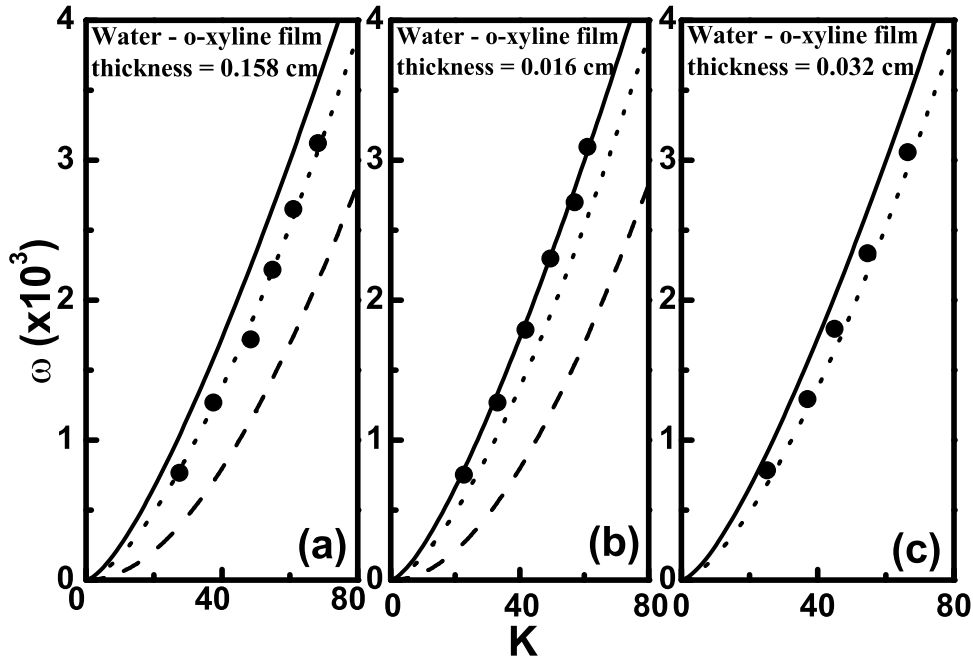}} 
\caption{Theoretical plot of frequency ($\omega$) versus  wavevector ($K$)
for different modes in water-o-xylene liquid-liquid film at
   film thickness (a) 0.158 cm, (b) 0.016 cm and (c) 0.032 cm. The solid line is for $\omega_{+}$ mode,
   dashed line is for $\omega_{-}$ mode and dotted line is for $\omega_{s}$
   mode. The experimental data points are shown
   by the filled  circles.}
\label{wox}
\end{figure}

In our experiments, we have tried to figure out which of
the two different modes (discussed
in previous section) that arise due to the presence of films of
kerosene, n,n-dimethyl aniline,or o-xylene spread on the surface of
water, is actually present. The choice of these liquids is due to 
the fact that these
are immiscible with water, have density lower w.r.t. water, are
low-viscous and have a higher boiling point. In these experiments
with films, a petridish of area $\sim$ 63 cm${^2}$  is filled with
water to a depth of nearly 1 cm and the liquid film is formed by
adding a measured quantity of the second liquid in drops from a
micropipette. The thickness of the film is estimated from the
known volume of the liquid and the surface area of the petridish.
Uniformity of the films on the water surface has been checked
optically. In order to avoid any decrease in thickness due to
evaporation, we have taken observations within 5 minutes of the
formation of the uniformly spread film. The measurements are of
the same type as those mentioned while studying surface tension
(discussed in the earlier section) except that, in the present
case, we do the experiments for varying film thicknesses. Here, we
show our experimental results only for o-xylene films of different
thicknesses on water. For films of the other liquids mentioned
above, on water, we have observed similar behavior.

To find out the mode which is present on the film surface for a liquid-film
of o-xylene on water, we first plot the experimental $\omega$
$\sim$ $K$ behavior with film thickness as a parameter. The
theoretically obtained $\omega$-$K$ curves for different modes :
$\omega_{+}$, $\omega_{-}$, and $\omega_{s}$ (as a limiting case
of $\omega_{+}$ mode) - are then placed on the same graph for
comparison. Here we use standard values : $\alpha$ = 37.2 dyne/cm,
$\alpha^{\prime}$ = 29.8 dyne/cm, $\rho^{\prime}$= 0.88 gm/cc,
 and $\rho$=1 gm/cc in Eqn. 22 \cite{Lide:1998}. It is evident that for a
film of
thickness as low as 0.016 cm, the measured data corresponds to
$\omega_{+}$ mode [Fig. \ref{wox}(b)], while for films of thickness
typically 0.158 cm, the experimental data matches well with the
$\omega_{s}$ mode [Fig. \ref{wox}(a)]. However, any intermediate
thickness between 0.016 and 0.158 cm yields data within the
$\omega_{+}$ and $\omega_{s}$ modes [Fig. \ref{wox}(c)]. This clearly
shows that the $\omega_{+}$ mode always dominates over
$\omega_{-}$ mode. For thicker films
the $\omega_+$ mode coincides with $\omega_s$ mode, whereas for
thinner films the mode is still $\omega_+$, though it is modified
by the presence of the interfacial tension and the density of the
lower liquid and is not equal to $\omega_s$.

\begin{figure}[htbp]
\centerline{\epsfxsize=4.25in\epsffile{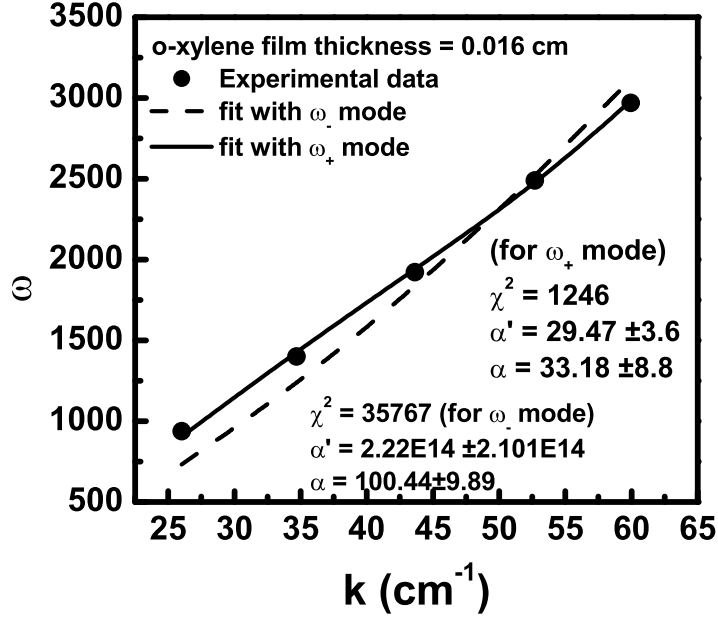}}
\caption{Nonlinear curve fitting of the experimental data for
water-
 o-xylene film of thickness 0.016 cm with $\omega_{+}$ and
$\omega_{-}$ modes.}
\label{fitting}
\end{figure}

The above conclusions  have been further verified by analyzing the
experimental data for the estimation of the surface and
interfacial tension. We have fitted our experimental data of
water-o-xylene film of thickness 0.016 cm with $\omega_{+}$ and
$\omega_{-}$ modes in Eqn. 22 by nonlinear curve fitting (keeping
surface tension and interfacial tension as varying parameters).
The  $\omega_{+}$ mode fits with the experimental data and
estimates the values of surface tension ($\alpha^\prime$) of
o-xylene as ($29.5 \pm 3.6 $) dyne/cm and the values of
interfacial tension ($\alpha$) as ($33.2 \pm 8.8$) dyne/cm [solid
line in Fig. \ref{fitting}].  These values are very close to the
standard values available in the literature \cite{Lord:1997}. On
the other hand, $\omega_{-}$ mode fails to fit the experimental
data well (dashed line in Fig. \ref{fitting}); moreover, it yields
unphysical values of surface and interfacial tension.
Experimental limitations, in our present set up, did not allow us
to make films of thickness lower than the values quoted above.

\section{Concluding remarks}

In this article, we describe a simple optics-based experimental
technique devised with the necessary background theoretical
formulation to study the characteristics of surface
capillary waves on liquids. Firstly, we visualize the profile of
the surface waves using simulations with realistic values for the
parameters involved. The locations of the nodes and antinodes
which appear in the simulated profile are used to match the
profile with actual experiments. We also claim that this combined
method of simulation and experimentation can be used for other
geometric configurations of oscillating sources on liquid
surfaces. Subsequently, the two key liquid properties, surface
tension and viscosity, which appear in the dispersion relation and
play an integral role in the propagation of such capillary waves
on a liquid surface, are measured optically using Fraunhofer
diffraction of laser light. We note, in particular, the novelty
and simplicity in the measurement of viscosity and the fact that
our measurement tools, are non-destructive but by no means, less
precise. Results of our experiments with several liquids, when
compared to the respective values known from other sources and
literature, summarily establish the efficacy and accuracy of our
approach. We anticipate, in future, the introduction of
technological sophistications in our set-up, which might lead to
an `Optical Surfacetensometer cum Viscometer'. The advantage of
having a single set-up for measuring both these properties need
not be further emphasised. Finally, we investigate, both
experimentally and theoretically, an interesting aspect of capillary wave
characteristics involved with a liquid film placed on the surface
of an immissible liquid. The dispersion relation we frame to study
this case not only explains the features of the
surface and interfacial modes, but also provides an
estimate of the interfacial tension across the film-liquid
boundary. In summary, our experimental and theoretical results
seem to demonstrate the capability of our set-up in carrying out
studies pertaining to capillary wave profile, liquid properties and
waves on liquid films on liquids. Despite its limitations,
the wide variety of information which can be obtained
from such a simple set-up, makes it a tool
worth improving upon in future investigations.

\section{Acknowledgment}

Anushree Roy and Tarun Kumar Barik thank Department of Science and
Technology (DST), New Delhi, India, for financial support. TKB
also thanks Dibyendu Chowdhury, an undergraduate student of
Vidyasagar University, West Bengal, India, for his help while
doing some of the experiments.

\newpage
\begin{table}
\begin{tabular}{|c|c|c|c|c|c|c|c|}\hline
{\bf Descriptions} & \multicolumn{6}{c}{\bf Angle between the
lines} &
 \\ \cline{2-8}  & { \bf A - B } &  { \bf B - C } & { \bf C - D } &
{ \bf D -
 E } & { \bf E - F } & { \bf F - G } & { \bf G - H }\\\hline
{\bf Simulated angular differences} &  &  &  &  &  &  &  \\
{\bf between the two corresponding} & \bf $ 25.5^{0}$ & $18^{0}$ &
$16^{0}$ & $15^{0}$ & $16^{0}$ & $18^{0}$ & $25.5^{0}$ \\
{\bf minima} &  &  &  &  &  &  & \\\hline
{\bf Experimental angular differences} &  &  &  &  &  &  &  \\
{\bf between the two corresponding} & \bf $ 25.1^{0}$ & $17.6^{0}$
&
$16^{0}$ & $15^{0}$ & $15.7^{0}$ & $17.5^{0}$ & $25.3^{0}$ \\
{\bf minima} &  &  &  &  &  &  & \\\hline
\end{tabular}
\caption{The simulated and experimental angular separations
between
  the two successive destructive minima of two source interference
  pattern [see Fig. 2(b) and Fig. \ref{schem}].}
\end{table}

\begin{table}[h]
\begin{tabular}{|c|c|c|c|c|c|}\hline
{\bf Liquids} & {\bf $\alpha$ in  dyne/cm } &  {\bf$ \alpha$ in
  dyne/cm} & {\bf$\delta$ in cgs } & {\bf$\eta$ in cp } & {\bf
  $\eta$ in
  cp}\\ & {\bf(experimental)} & {\bf(standard)} & {\bf units} &
  {\bf(experimental)} & {\bf(standard)}  \\\hline &  &  & {$(0.235 \pm
  0.026)$}  & {$(0.92 \pm 0.1)$ at 220 Hz} & \\
{\bf Water} & 75 $\pm$ 5.3  & 72 &{$(0.290 \pm
  0.024)$}  & {$(0.96 \pm 0.08)$ at 260 Hz} &
  0.89  \\ &  &  &{$(0.304 \pm
  0.031)$} & {$(0.87 \pm 0.09)$ at 300 Hz} & \\\hline &  &  &{$(0.947 \pm
  0.04)$} &
{$(1.44 \pm 0.06)$ at 220 Hz} & \\
{\bf Kerosene} & 26.3 $\pm$ 3.7 & 28 & {$(1.11 \pm
  0.05)$} & {$(1.43 \pm 0.07)$ at 260 Hz} &
  1.40  \\ &  &  &  &  & \\\hline
{\bf 1-Hexanol} & 28.5 $\pm$ 3.4 & 26 & {$(2.75 \pm
  0.17)$} & {$(4.27 \pm 0.26)$ at 220
  Hz} & 4.58 \\\hline
\end{tabular}
\caption{The experimental and standard values of surface tension
  ($\alpha$) and viscosity ($\eta$) of different liquids at room
  temperature ($25^{0}$ C). For standard values of $\alpha$ and $\eta$
  see reference \cite{Lide:1998}.}
\end{table}


\end{document}